\documentclass[a4paper]{article}

\usepackage{INTERSPEECH2022}
\usepackage{comment}
\usepackage{url}
\usepackage{hyperref}
\usepackage{amsmath,amssymb}
\usepackage{bbm}
\usepackage{subfigure}
\hypersetup{
    colorlinks = false,
}
\usepackage{cite}
\title{UTMOS: UTokyo-SaruLab System for VoiceMOS Challenge 2022}
\name{Takaaki Saeki$^{*}$, Detai Xin$^{*}$, Wataru Nakata$^{*}$,\\Tomoki Koriyama, Shinnosuke Takamichi, and Hiroshi Saruwatari\thanks{$^*$Equal contribution.}}
\address{The University of Tokyo, Japan.}
\email{\{takaaki\_saeki, detai\_xin\}@ipc.i.u-tokyo.ac.jp, nakata-wataru855@g.ecc.u-tokyo.ac.jp}

\begin{document}
\setlength{\abovedisplayskip}{3pt} 
\setlength{\belowdisplayskip}{3pt} 
\setlength\floatsep{3pt} 
\setlength\intextsep{5pt} 
\setlength\textfloatsep{10pt} 
\setlength\abovecaptionskip{5pt} 
\setlength{\dbltextfloatsep}{6pt} 

\maketitle
\begin{abstract}\vspace{-2mm}
We present the UTokyo-SaruLab mean opinion score (MOS) prediction system submitted to VoiceMOS Challenge 2022.
The challenge is to predict the MOS values of speech samples collected from previous Blizzard Challenges and Voice Conversion Challenges for two tracks: a main track for in-domain prediction and an out-of-domain (OOD) track for which there is less labeled data from different listening tests.
Our system is based on ensemble learning of strong and weak learners.
Strong learners incorporate several improvements to the previous fine-tuning models of self-supervised learning (SSL) models, while weak learners use basic machine-learning methods to predict scores from SSL features.
In the Challenge, our system had the highest score on several metrics for both the main and OOD tracks.
In addition, we conducted ablation studies to investigate the effectiveness of our proposed methods.

\end{abstract}
\noindent\textbf{Index Terms}: VoiceMOS Challenge 2022, mean opinion score prediction, self-supervised learning, ensemble learning

\vspace{-2mm}
\section{Introduction}\label{sec:intro}
\vspace{-1mm}

Although subjective evaluation has been the gold standard in the field of speech synthesis~\cite{black05blizzard}, its high cost in terms of time and money motivates us to develop measures for automatically determining the performance.
Although a number of neural network-based approaches for doing this have been proposed~\cite{patton16automos,lo2019mosnet,Leng2021MBNETMP}, there are still many challenges, such as developing a general-purpose prediction model.

The VoiceMOS Challenge~\cite{huang21voicemos}, which was launched this year, provides the common database and baseline systems.
The database contains synthetic speech samples and the corresponding mean opinion scores (MOS) on a five-point scale as assigned by human evaluators.
The participants construct a prediction system and submit the system's predicted MOS for the test data.
There are two tracks in the challenge, the main and out-of-domain (OOD) tracks, and the system performance is evaluated using several metrics.


In this paper, we present our MOS prediction system, UTMOS (pronounced ``u--t--mos''), which we submitted to VoiceMOS Challenge 2022.
Our system is based on ensemble learning of strong and weak learners: the strong learners are obtained by fine-tuning models of self-supervised learning (SSL) models, and the weak learners predict scores from SSL features by using non-neural-network machine-learning methods from SSL features.
The strong learner incorporates several improvement functions, including contrastive learning, listener dependency, and phoneme encoding.
We also present the results of VoiceMOS Challenge 2022 and those of our ablation studies.
Our implementation is publicly available\footnote{\url{https://github.com/sarulab-speech/UTMOS22}}.
This paper makes three contributions in particular:
\vspace{-1mm}
\begin{itemize} \leftskip -5.5mm \itemsep -0.5mm
    \item It describes and MOS prediction system that had the highest score on several metrics in the main and OOD tracks of VoiceMOS Challenge 2022.
    \item It presents proposed methods for predicting MOS that include contrastive learning and phoneme encoding.
    \item It presents the results of ablation studies demonstrating the effectiveness of listener-dependent learning and that of stacking by increasing the number of strong learners.
\end{itemize}

\vspace{-2mm}
\section{VoiceMOS Challenge 2022}\label{sec:voicemos}
\vspace{-1mm}

The data used in the VoiceMOS Challenge 2022 were mainly synthetic speech samples from previous Blizzard Challenges and Voice Conversion Challenges.
The VoiceMOS Challenge is divided into two tracks: the main track and the OOD track.
The dataset statistics for both tracks are given in Table~\ref{tab:dataset}.

\textbf{Main track.}
The main track uses the BVCC dataset~\cite{cooper2021bvcc}, which contains data obtained by conducting large-scale listening tests on samples from 187 different systems from previous Blizzard Challenges, Voice Conversion Challenges, and ESPnet-TTS~\cite{Hayashi2020EspnetTTSUR} published samples.
The main track dataset consists of English synthetic speech samples.
The test set contains samples obtained with the same listening test but from unseen systems, speakers, and listeners.

\textbf{OOD track.}
The OOD track uses a dataset consisting of Chinese synthetic speech collected using a listening test different from that used for the main track.
The dataset provides a small amount of labeled data and a large amount of audio-only unlabeled data.

For both the main and OOD tracks, prediction performance was evaluated using four metrics; mean squared error (MSE), linear correlation coefficient (LCC), Spearman rank correlation coefficient (SRCC), and Kendall rank correlation coefficient (KTAU).
Participants submit their predicted score for each speech utterance in the test set, and utterance-level and system-level metrics were calculated for each of the four metrics.

\begin{table*}[tb]
\centering
\caption{Datasets used in VoiceMOS Challenge 2022: ``closed/open'' indicates that system used to synthesize speech is included/excluded in training data; ``labeled/unlabeled'' indicates that corresponding MOS score is included/excluded.}
\label{tab:dataset}
\vspace{-1mm}
\scalebox{0.68}{
\begin{tabular}{l|ccc|ccc}
\toprule
                       & \multicolumn{3}{c|}{Main}                           & \multicolumn{3}{c}{OOD}                            \\
                       & Train              & Dev                     & Test & Train            & Dev                      & Test \\ \midrule
Evaluations             & 39,792              & 8,258                    &  8,528    & 1,848             & 1,819                     &  7,680    \\
Audio clips            & 4,974               & 1,066 (open)             &  1,066 (open)    & 136 (labelled) + 540 (unlabelled)             & 136 (open)               &  540 (open)    \\
Audio clips per system & 12--36 (avg: 29.4) & 1--37 (avg: 5.9)        &  1--38 (avg: 5.7)    & 4--10 (avg: 6.5) & 4--46 (avg: 2.5)         & 6--52 (avg: 20.8)      \\
Systems                & 175                & 175 (closed) + 6 (open) &  175 (closed) + 12 (open)    & 21               & 21 (closed) + 3 (open)   & 21 (closed) + 5 (open)    \\
Listeners               & 288                & 288 (closed) + 8 (open) & 288 (closed) + 16 (open)     & 285              & 285 (closed) + 43 (open) &  285 (closed) + 76 (open)    \\ \bottomrule
\end{tabular}
}
\end{table*}

\vspace{-2mm}
\section{UTMOS}\label{sec:method}
\vspace{-1mm}

\begin{figure}[t]
  \centering
  \includegraphics[width=1.0\linewidth, clip]{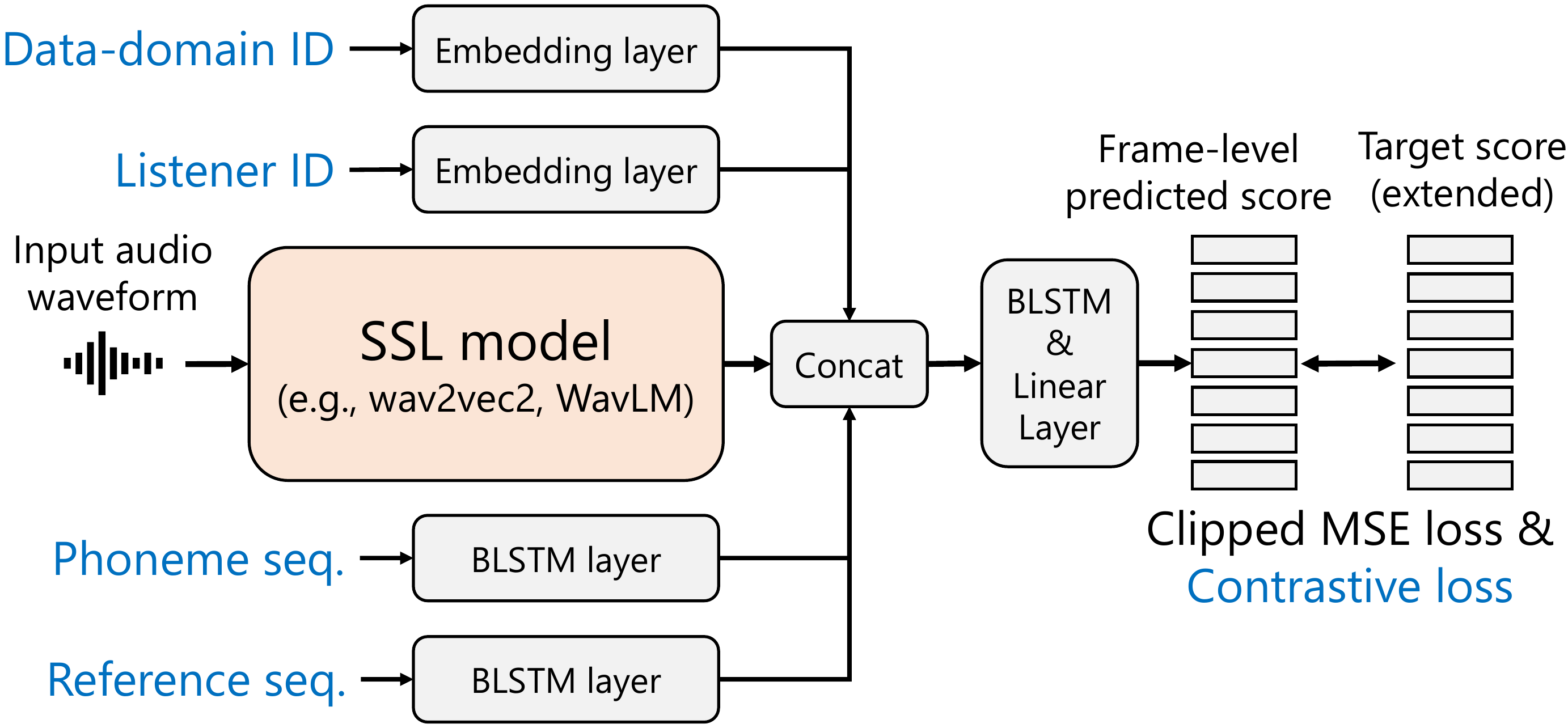}
  \caption{Architecture of the proposed strong learner.}
  \label{fig:strong}
\end{figure}

Our UTMOS method leverages ensemble learning by using multiple models, which consists of \textit{strong} learners and \textit{weak} learners. The strong learner is an SSL-based neural network model that directly uses a speech waveform as input. The weak learners are basic machine learning models, such as ridge regression models and support vector machines with utterance-level SSL features as input.

\vspace{-2mm}
\subsection{Fine-tuned SSL model}\label{sec:method-strong}
\vspace{-1mm}

\subsubsection{Basic architecture}\label{sec:method-strong-basic}
\vspace{-2mm}

Fig.~\ref{fig:strong} illustrates the architecture of the strong learner.
As in a previous study~\cite{Cooper2021GeneralizationAO}, we used a pretrained SSL model to extract features from input audio.
First, the raw waveform of a speech utterance is input to the SSL model to obtain frame-level features.
Unlike the previous study~\cite{Cooper2021GeneralizationAO}, our model does not average the frame-level output features; instead, it sends them to bidirectional long sort-term memory (BLSTM) and linear layers to compute frame-level scores.
During training, we extend the target score by the number of frames and define a frame-level loss function.
We found that the frame-level loss achieves higher performance than the previous one using averaged features.
During inference, the model predicts the score by averaging the frame-level scores.
Using this model, we introduced several functions described in Sections~\ref{sec:method-contrastive} to \ref{sec:method-aug} for the strong learner.

\vspace{-2mm}
\subsubsection{Contrastive loss}\label{sec:method-contrastive} \vspace{-2mm}
Contrastive learning is a self-supervised machine-learning method that can utilize unlabeled data by learning from intrinsic similarity relations between data.
Contrastive learning is widely used in speech quality assessment, in which speech representations are learned from large-scale unlabeled speech data \cite{serra2021sesqa, manocha2021cdpam, manocha2021noresqa}.
Given scores $s_{1}$ and $s_{2}$ of utterances $x_{1}$ and $x_{2}$, the difference in the scores ($d_{x_{1}, x_{2}} = s_{1} - s_{2}$) can be regarded as the difference in the two utterances in terms of speech quality.
If the predicted scores for the two utterances are denoted as $\widehat{s}_{1}$ and $\widehat{s}_{2}$, respectively, it is intuitive to assume that the predicted difference ($\widehat{d}_{x_{1}, x_{2}} = \widehat{s}_{1} - \widehat{s}_{2}$) is close to $d_{x_{1}, x_{2}}$.
Therefore, we consider a contrastive loss defined as $\mathcal{L}^{\mathrm{con}}_{x_{1}, x_{2}} = \max(0, | d_{x_{1}, x_{2}} - \widehat{d}_{x_{1}, x_{2}} | - \alpha)$,
where $\alpha$ is a hyperparameter greater than zero.
We call $\alpha$ the margin since it is similar to the support vector machine margin, i.e. small errors are ignored by the model.
One advantage of this loss function is that it penalizes the model when the signs of $d_{x_{1}, x_{2}}$ and $\widehat{d}_{x_{1}, x_{2}}$ are opposite, e.g. the case in which $x_{1}$ is better than $x_{2}$ but the model predicts that $x_{1}$ is worse than $x_{2}$, which makes the contrastive loss suitable for improving the metrics based on ranking accuracy, such as the SRCC used in the Challenge.

In practice the contrastive loss of all possible pairs in a mini-batch are considered: $\mathcal{L}^{\mathrm{con}} = \textstyle \sum_{i \neq j} \mathcal{L}^{\mathrm{con}}_{x_{i}, x_{j}}$.
It is worth noting that, as discussed in Section~\ref{sec:ablation}, our model can be trained using only the proposed contrastive loss function without using other regular loss functions like MSE or L1 and still achieve better performance than the baseline model.

In addition to the contrastive loss, we use the clipped MSE loss~\cite{Leng2021MBNETMP} for the regression loss between the discrete predicted and ground truth scores: $\mathcal{L}^{\mathrm{reg}}(y, \hat{y}) = \mathbbm{1} (|y - \hat{y}| > \tau)(y-\hat{y})^2$.
The final loss function is defined as:
\begin{equation}
    \mathcal{L} = \beta \mathcal{L}^{\mathrm{reg}} + \gamma \mathcal{L}^{\mathrm{con}} \label{eq:loss}
\end{equation}
where $\beta$ and $\gamma$ are hyperparameters.

\vspace{-2mm}
\subsubsection{Data-domain and listener dependent learning}\label{sec:method-strong-ld} \vspace{-2mm}
The previous MOS prediction model based on an SSL model~\cite{Cooper2021GeneralizationAO} learns the utterance-level MOS as the target variable.
Previous studies~\cite{Leng2021MBNETMP,huang21ldnet} improved the prediction accuracy by making listener-dependent predictions instead of simply predicting utterance-level MOS; accuracy was improved because the distribution of evaluation scores is different for each listener.

We thus introduced a listener dependency function into the SSL-based MOS predictor.
As shown in Fig.~\ref{fig:strong}, listener-embedding is concatenated with the features extracted by the SSL model to predict the listener-dependent score.
During training, we also include a ``mean listener'' for which the target score is the average value of all the listeners' scores, as in a previous study~\cite{huang21ldnet}
Since the listener information is unknown in inference, the mean-listener embedding is given to the model to predict the utterance-level MOS.

Furthermore, we need to consider the bias of each listening test instead of considering only the bias per listener within a given listening test.
For example, different listening tests are conducted for the main and OOD datasets.
To include data from multiple domains in our training, we use both listener and domain IDs as shown in Fig.~\ref{fig:strong}.
When we train our model on all of the main, OOD, and external datasets described in Section~\ref{sec:method-external}, for example, we assign different domain IDs to the respective datasets.
In addition, we use the average score within each domain to obtain the score of the mean listener.

\vspace{-2mm}
\subsubsection{Phoneme encoding}\label{sec:method-phoneme} \vspace{-2mm}
In our preliminary studies, we observed that there was a strong correlation between the MOS ratings and clustering results of linguistic contents estimated with an automatic speech recognition (ASR) model.
On the basis of this observation, we use the ASR results as auxiliary input of the strong learner to further improve prediction accuracy.
To apply this method to multilingual synthetic speech samples, we use phoneme sequences as input instead of graphemes or character sequences.  

Furthermore, intuitively, the larger the difference between the reference text used to generate synthetic speech and the text estimated with ASR, the lower the intelligibility and the expected MOS rating.
Since the participants were not provided the actual texts used for synthesized speech, to estimate the reference text of each utterance, we perform clustering on the ASR results by using the DBSCAN algorithm~\cite{dbscan} based on normalized Levenshtein distance and extract the median text corresponding to each cluster. We refer to this median text as the reference sequence.
As shown in Fig.~\ref{fig:strong}, phoneme and reference sequences are fed to the phoneme encoder which consists of BLSTM layers and the initial and last hidden states are concatenated.
Finally it is replicated with the number of frames and concatenated to the output of the SSL model.

\vspace{-2mm}
\subsubsection{Data augmentation}\label{sec:method-aug} \vspace{-2mm}
Deep neural networks usually suffer from overfitting if the training data is limited.
Given the limited data size of the challenge, especially for the OOD track, overfitting is likely to happen.
We thus utilize data augmentation to alleviate this problem.
We consider two augmentation methods: speaking-rate-changing and pitch-shifting, which can alter the utterances while maintaining the MOS.
Speaking-rate-changing slows down or speeds up the audio by a factor $f_{t}$.
Since a very large or small $f_{t}$ will affect the MOS, we set $f_{t}$ close to $1$.
Pitch-shifting changes the speaker identity of the utterances by raising or lowering the original pitch $p$ to $p + f_{p}$.
During training, the two parameters ($f_{t}$ and $f_{p}$) are randomly selected from two ranges $[1 - F_{t}, 1 + F_{t}]$ and $[-F_{p}, F_{p}]$, respectively.
We tune $F_{p}$ so that the MOS of the augmented waveform has perceptually little difference from the original ones.
We use WavAugment~\cite{wavaugment2020} to implement all data augmentation methods.

\vspace{-2mm}
\subsection{External data collection}\label{sec:method-external} \vspace{-1mm}
In the OOD track, the size of the labeled training data (136 utterances) is not sufficient to train a robust MOS prediction model.
Therefore we utilized the 540 unlabeled utterances and collecting the corresponding MOS as external data.
To this end, we first selected the system with the highest MOS (BC2019-A) and regarded all utterances of this system as natural speech after double-checking the utterances with a Chinese native speaker.
We then conducted a standard 5-point-scale MOS test for all 540 unlabeled utterances and 249 labeled utterances.
A total of 32 Chinese listeners participated in this test; each listener rated 55 utterances, so each utterance had 2 answers on average.
The utterance-level SRCC between the ground truth scores and collected scores for the labeled data was $0.757$, which indicates a strong correlation.
Although the distribution of these collected scores was not exactly the same as that of the original utterances, since they were all evaluated by Chinese, we think it is appropriate to utilize the external data along with the original data.
Using the external data substantially improved performance, as discussed in Section~\ref{sec:eval}.

\vspace{-2mm}
\subsection{Ensemble learning with strong and weak learners}\label{sec:method-stacking} \vspace{-1mm}

\begin{figure}[t]
  \centering
  \includegraphics[width=1.0\linewidth, clip]{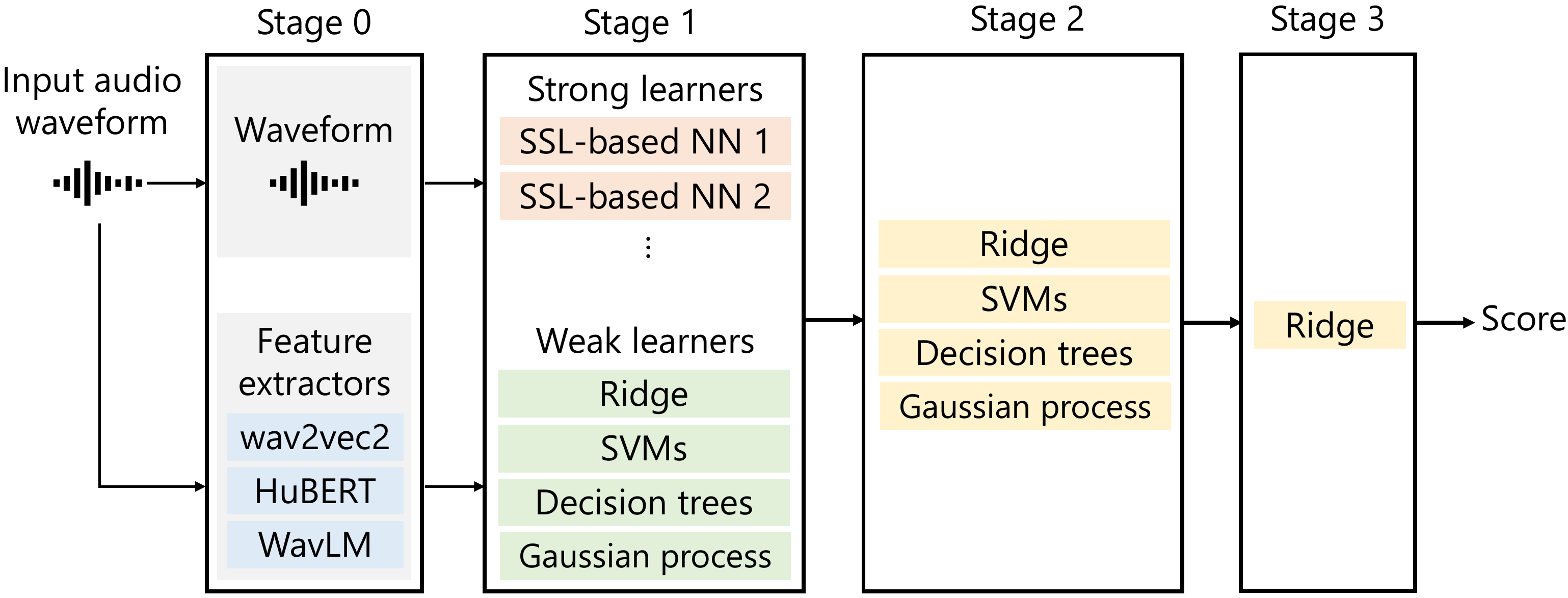}
  \caption{Flow of stacking with strong and weak learners.}
  \label{fig:stacking}
\end{figure}

We use an ensemble of models for prediction robustness. Specifically, we use the stacking method \cite{wolpert1992stacked,breiman1996stacked} illustrated in Fig.~\ref{fig:stacking} as the ensemble method. We use not only fine-tuned SSL models but also simple regression models using utterance-level features.
We refer to the former and latter models as ``strong learners'' and ``weak learners,'' respectively.

The weak learners are a combination of feature extractions and regression methods. We propose using pretrained-SSL-model-based mean embeddings for feature extraction. Specifically, we extract embeddings of input utterances and compute the mean for all frames, taking as inspiration the structure of SSL-MOS \cite{Cooper2021GeneralizationAO}.
Although this process might be too rough for obtaining utterance-level characteristics, we assume that only the mean embeddings have efficient information for MOS prediction.
For the simple regression models, we use basic ones such as linear regression, decision-tree-based methods, and kernel methods.
In general, model diversity is important for prediction performance in ensemble training \cite{zhou2012ensemble}.
Hence, we use multiple pretrained SSL models for feature extraction to increase the number of models.
Moreover, we enhance the diversity of weak learners by using different data domains, i.e. languages and MOS test environments, for the OOD track.

The stacking method comprises stages 0 to 3. After feature extraction, we train strong and weak learners individually and predict scores using cross validation. We then train meta learners using the first stage scores.
Finally, we train the third stage model with the second stage scores and obtain the final score.

\vspace{-2mm}
\section{Experimental evaluations}\label{sec:eval}
\vspace{-1mm}

\begin{table}[tb]
    \centering
    \setlength{\tabcolsep}{1mm} 
    \caption{Results of ablation study.}
    \vspace{-2mm}
    \label{tab:eval-ablation}
    \subtable[\textbf{Main}]{
    \scalebox{0.7}{
    \begin{tabular}{l|cccccccc}
    \toprule
                          & \multicolumn{4}{c}{Utterance-level}                                                    & \multicolumn{4}{c}{System-level}                                  \\
                          & MSE            & LCC            & SRCC           & \multicolumn{1}{c|}{KTAU}           & MSE            & LCC            & SRCC           & KTAU           \\ \midrule
    UTMOS strong          & 0.276          & 0.883          & 0.881          & \multicolumn{1}{c|}{0.708}          & 0.148          & 0.930          & 0.925          & 0.774          \\
    w/o contrastive loss  & 0.241          & 0.881          & 0.879          & \multicolumn{1}{c|}{0.706}          & 0.114          & 0.932          & 0.930          & 0.781          \\
    w/o listener ID       & \underline{0.307}    & \underline{0.880}    & \underline{0.878}    & \multicolumn{1}{c|}{\underline{0.704}}    & \underline{0.160}    & 0.935          & 0.933          & 0.784          \\
    w/o phoneme encoder   & 0.249          & 0.881          & 0.882          & \multicolumn{1}{c|}{0.709}          & 0.119          & 0.935          & \textbf{0.936} & \textbf{0.790} \\
    w/o data augmentation & 0.226          & \textbf{0.885} & \textbf{0.882} & \multicolumn{1}{c|}{\textbf{0.710}} & \textbf{0.103} & \textbf{0.936} & 0.933          & 0.784          \\
    w/o MSE loss          & \textbf{0.219} & 0.882          & 0.880          & \multicolumn{1}{c|}{0.707}          & 0.114          & 0.932          & 0.929          & 0.778          \\ \midrule
    SSL-MOS               & 0.380          & 0.869          & 0.871          & \multicolumn{1}{c|}{0.695}          & 0.223          & 0.920          & 0.918          & 0.758          \\ \bottomrule
    \end{tabular}
    }
    }
    \subtable[\textbf{OOD}]{
    \vspace{-3mm}
    \scalebox{0.7}{
    \begin{tabular}{l|cccccccc}
    \toprule
                          & \multicolumn{4}{c}{Utterance-level}                                                    & \multicolumn{4}{c}{System-level}                                  \\
                          & MSE            & LCC            & SRCC           & \multicolumn{1}{c|}{KTAU}           & MSE            & LCC            & SRCC           & KTAU           \\ \midrule
    UTMOS strong          & 0.378          & 0.891          & 0.871          & \multicolumn{1}{c|}{0.690}          & 0.248          & \textbf{0.970} & \textbf{0.972} & \textbf{0.879} \\
    w/o contrastive loss  & 0.407          & 0.870          & 0.862          & \multicolumn{1}{c|}{0.676}          & 0.272          & 0.945          & 0.957          & 0.841          \\
    w/o listener ID       & \underline{0.636}    & \underline{0.847}    & \underline{0.825}    & \multicolumn{1}{c|}{\underline{0.638}}    & \underline{0.490}    & \underline{0.931}    & \underline{0.944}    & \underline{0.820}    \\
    w/o phoneme encoder   & 0.390          & \textbf{0.893} & \textbf{0.881} & \multicolumn{1}{c|}{\textbf{0.702}} & 0.258          & 0.966          & 0.967          & 0.868          \\
    w/o data augmentation & \textbf{0.322} & 0.887          & 0.872          & \multicolumn{1}{c|}{0.691}          & \textbf{0.191} & 0.960          & 0.967          & 0.872          \\
    w/o external data     & 0.412          & 0.883          & 0.868          & \multicolumn{1}{c|}{0.684}          & 0.253          & 0.960          & 0.961          & 0.861          \\ \midrule
    SSL-MOS               & 0.676          & 0.872          & 0.842          & \multicolumn{1}{c|}{0.654}          & 0.500          & 0.957          & 0.964          & 0.862          \\ \bottomrule
    \end{tabular}
    }
    }
    \vspace{-2mm}
\end{table}

\begin{table}[tb]
    \centering
    \setlength{\tabcolsep}{1mm} 
    \caption{Results of staking. ``Strong'' and ``Weak'' are the number of strong and weak learners used for stacking, except the case that Strong is 1 and Weak is 0, which means a single SSL model is used. With regard to the numbers of weak learners at OOD track, 48, 96, and 144 corresponds to 1 (OOD), 2 (OOD, external), and 3 (OOD, external, main) domains, respectively.}
    \vspace{-2mm}
    \label{tab:eval-stacking}
    \subtable[\textbf{Main}]{
    \scalebox{0.70}{
    \begin{tabular}{cc|cccccccc}
    \toprule
           &      & \multicolumn{4}{c}{Utterance-level}                                                    & \multicolumn{4}{c}{System-level}                                  \\
    Strong & Weak & MSE            & LCC            & SRCC           & \multicolumn{1}{c|}{KTAU}           & MSE            & LCC            & SRCC           & KTAU           \\ \midrule
    1      & 0    & 0.216          & 0.894          & 0.890          & \multicolumn{1}{c|}{0.720}          & 0.105          & 0.937          & 0.934          & 0.792          \\
    17     & 0    & 0.169          & 0.896          & 0.893          & \multicolumn{1}{c|}{0.725}          & \textbf{0.088} & \textbf{0.939} & \textbf{0.936} & 0.792          \\ \midrule
    0      & 48   & 0.186          & 0.887          & 0.885          & \multicolumn{1}{c|}{0.714}          & 0.108          & 0.928          & 0.927          & 0.777          \\ \midrule
    1      & 48   & 0.172          & 0.896          & 0.894          & \multicolumn{1}{c|}{0.726}          & 0.098          & 0.935          & 0.933          & 0.789          \\
    5      & 48   & 0.169          & 0.898          & 0.895          & \multicolumn{1}{c|}{0.728}          & 0.095          & 0.938          & \textbf{0.936} & 0.793          \\
    12     & 48   & 0.169          & 0.898          & 0.895          & \multicolumn{1}{c|}{0.728}          & 0.094          & 0.938          & 0.935          & 0.792          \\
    17     & 48   & \textbf{0.165} & \textbf{0.899} & \textbf{0.896} & \multicolumn{1}{c|}{\textbf{0.730}} & 0.090          & \textbf{0.939} & \textbf{0.936} & \textbf{0.795} \\ \bottomrule
    \end{tabular}
    }
    }
    \subtable[\textbf{OOD}]{
    \vspace{-3mm}
    \scalebox{0.70}{
    \begin{tabular}{cc|cccccccc}
    \toprule
        &      & \multicolumn{4}{c}{Utterance-level}                                                    & \multicolumn{4}{c}{System-level}                                  \\
    Strong & Weak & MSE            & LCC            & SRCC           & \multicolumn{1}{c|}{KTAU}           & MSE            & LCC            & SRCC           & KTAU           \\ \midrule
    1     & -    & 0.280          & 0.905          & 0.885          & \multicolumn{1}{c|}{0.704}          & 0.160          & 0.972          & 0.965          & 0.858          \\
    6      & 0    & \textbf{0.155} & \textbf{0.920} & \textbf{0.896} & \multicolumn{1}{c|}{\textbf{0.720}} & 0.029          & 0.988          & 0.975          & 0.886          \\ \midrule
    0      & 48   & 0.204          & 0.893          & 0.858          & \multicolumn{1}{c|}{0.674}          & 0.033          & 0.985          & 0.963          & 0.860          \\
    0      & 96   & 0.179          & 0.907          & 0.877          & \multicolumn{1}{c|}{0.696}          & 0.030          & 0.988          & 0.975          & 0.890          \\
    0      & 144  & 0.176          & 0.909          & 0.882          & \multicolumn{1}{c|}{0.702}          & 0.033          & 0.987          & 0.974          & 0.888          \\ \midrule
    1      & 144  & 0.174          & 0.910          & 0.883          & \multicolumn{1}{c|}{0.704}          & 0.033          & 0.986          & 0.976          & 0.894          \\
    6      & 144  & 0.162          & 0.917          & 0.892          & \multicolumn{1}{c|}{0.715}          & \textbf{0.028} & \textbf{0.989} & \textbf{0.977} & \textbf{0.900} \\ \bottomrule
    \end{tabular}
    }
    }
    \vspace{-2mm}
\end{table}

\subsection{Experimental conditions} \vspace{-1mm}
For the strong learners, we trained models with six different configurations for the main track and six different configurations for the OOD track.
The main configuration was \textbf{UTMOS strong}, the strong learner used for submission. We also trained the strong learner without the functions described from Sections~\ref{sec:method-contrastive} to \ref{sec:method-external} and the use of MSE loss for the main track and the use of external data for the OOD track.
For the preprocessing, we downsampled all speech samples to 16~kHz and normalized the volume. During training, each MOS rating was normalized to the range [-1, 1] by applying linear projection. 
For the SSL models of strong learners, we used the published wav2vec2.0~\cite{baevski2020wav2vec} base model\footnote{\url{https://github.com/pytorch/fairseq/blob/main/examples/wav2vec}} pretrained on Librispeech~\cite{librispeech}.
For phoneme transcription in the phoneme encoder, we used the ASR model proposed by Xu et. al\cite{xu2021simple}.
This model is xlsr-53~\cite{conneau21_interspeech} fine-tuned on phonetic annotations from word transcriptions obtained using ESpeak\footnote{\url{https://github.com/espeak-ng/espeak-ng}} and speech samples of CommonVoice~\cite{ardila-etal-2020-common}.
For the phoneme encoder, we used 3-layer BLSTM with a hidden size of 256.
For domain and listener embedding, we used an embedding dimension of 128.
For the main track, we only used main track dataset. 
For the OOD track, we used the OOD track dataset, and external dataset we collected for training except when trained for ``w/o external data''. For ``w/o external data'', we used only the dataset from OOD track for training.
For the hyperparameters of the loss function defined in Eq.~(\ref{eq:loss}), we set $\beta = 1$, and $\gamma = 0.5$ except for ``w/o contrastive loss'' and ``w/o MSE loss.'' For ``w/o contrastive loss'' and ``w/o MSE loss,'' we used  $\beta = 1, \gamma=0$, and  $\beta=0,\gamma=1$, respectively.
For $\alpha$ and $\tau$, we set $\alpha = 0.5, \tau=0.25$ except for ``w/o listener ID.'' For ``w/o listener ID,'' we set $\alpha=0.1, \tau=0.1$.
For data augmentation, we set $F_t = 0.1$, and $F_p = 300$ cents except for ``w/o data augmentation.'' For ``w/o data augmentation,'' no data augmentation was performed. 
For the optimizer, we used Adam~\cite{kingma14adam} ($\beta_1 = 0.9, \beta_2 = 0.99$) with linear warmup and linear decay learning rate scheduling.
Learning rate warmup was performed for 4000 steps, and the total number of training steps was 15,000.
The batch size was 12, and gradient accumulation was performed every 2 steps.
The best model checkpoint was selected on the basis of the highest system-level SRCC calculated from the development set.
For the ablation study of strong learners and stacking, training was performed for five times and the results were calculated by averaging scores for each metric as model performance varies depending on the random seed.

Regarding the conditions for stacking and weak learners,
the strong learners for stacking were chosen from the candidates during hyperparameter tuning by using the Optuna~\cite{akiba2019optuna} based on the system-level SRCC of development set.
We used a maximum of 17 and 6 strong learners for the main and OOD tracks, respectively.
For the pretrained SSL features for weak leaners,
we used four wav2vec 2.0 \cite{baevski2020wav2vec}, two HuBERT \cite{hsu2021hubert}, and two WavLM \cite{chen2022wavlm} models,
which differed from each other in model size, database, and training method.
The simple regression methods for the weak and meta learners were two linear regressions (ridge regression and linear support vector regression (SVR)),
two tree-based models (random forests and LightGBM \cite{guolin2017lightgbm}),
and two kernel methods (kernel SVR and Gaussian process regression).
By combining pretrained SSL models and simple regression methods, we obtained 48 weak learners.

The training of the weak and meta learners for the main track was performed using only the main track data.
For the OOD track, we trained weak models for respective domains, which were main, OOD, and external ones, and integrated the results at the second stage. Hence, the number of weak learners for the OOD track was 144. The meta learners for the OOD track were trained using the OOD track data.

\vspace{-2mm}
\subsection{VoiceMOS2022 results} \vspace{-1mm}
In the both tracks, utterance-level (Utt.) and system-level (Sys.) metrics were calculated as described in Seciton~\ref{sec:voicemos}.
Three baseline methods and the 21 teams participated in the Main track.
For the OOD track, scores of three baseline methods and 15 teams were submitted.
Our team ID is ``T17.''

Part of our results in the Main track were $\text{Utt. MSE} = 0.165 \, (1)$, $\text{Utt. SRCC} = 0.897 \, (1)$, $\text{Sys. MSE} = 0.090 \, (1)$, $\text{Sys. SRCC} = 0.936 \, (3)$, where the numbers in parentheses mean the rankings.
The results in the OOD track were $\text{Utt. MSE} = 0.162 \, (1)$, $\text{Utt. SRCC} = 0.893 \, (2)$, $\text{Sys. MSE} = 0.030 \, (1)$, $\text{Sys. SRCC} = 0.988 \, (1)$.

\vspace{-2mm}
\subsection{Ablation study on SSL-based models}
\label{sec:ablation}\vspace{-1mm}
We conducted ablation studies for each of the methods described in Section~\ref{sec:method-strong}.
We denote a strong learner using all the methods in Section~\ref{sec:method-strong} as ``UTMOS strong.''
A method based on fine-tuning of the SSL model~\cite{Cooper2021GeneralizationAO}, which is a baseline method of the challenge, was designated as ``SSL-MOS.''
Table~\ref{tab:eval-ablation} lists the results.
The best results are shown in bold, while the worst ones are underlined except for UTMOS strong and SSL-MOS.

We can see that all of our methods outperformed SSL-MOS in almost all indices.
Furthermore, in the main track, the method that excluded data augmentation or phoneme encoder from UTMOS strong showed better results, which may be due to the larger amount of data of the main track than the OOD track.
In the OOD track, UTMOS strong showed the best results in several indices including test system SRCC.
This suggests that all of the proposed methods have effectiveness in cases with smaller amounts of data.
For both the main and OOD tracks, the performance of the methods without the listener ID significantly degraded in many cases, indicating the effectiveness of listener dependency.

\vspace{-2mm}
\subsection{Evaluation on stacking} \vspace{-1mm}

To investigate the effectiveness of strong and weak learners at stacking, we computed the prediction accuracy scores by changing the number of strong and weak learners.
The results are shown in Table~\ref{tab:eval-stacking}.
The 1, 5, and 12 strong learners were chosen greedily based on system-level SRCC of development set.

We can see that even single strong layer gave high SRCCs
although the MSEs were still large. 
By using the stacking ensemble with multiple strong learners, MSEs were reduced while SRCCs were kept high.
The stacking using only weak models even had
high SRCCs
although the extracted features were simpler than fine-tuned SSL models.
We also see that the increase of the number of strong and weak learners tended to improve prediction accuracy, which indicates it is promising to increase the number of models by using multiple hyperparameter values and multiple domains.

\vspace{-2mm}
\section{Conclusion} \vspace{-1mm}
We presented the system we submitted to VoiceMOS Challenge 2022.
Our system is based on ensemble learning of strong learners, which are obtained by fine-tuning SSL models, and weak learners that predict scores from SSL features.
Future work includes constructing a larger-scale general-purpose MOS prediction model by collecting a wider variety of data.

{\footnotesize
\textbf{Acknowledgements:} 
Part of this work was supported by JSPS KAKENHI Grant Number 21H04900, 21K11955, and JST SPRING, Grant Number JPMJSP2108 (for implementation) and JST Moonshot R\&D Grant Number JPMJPS2011 (for evaluation).
}

\bibliographystyle{IEEEtran}
\bibliography{tts}

\begin{thebibliography}{10}
\providecommand{\url}[1]{#1}
\csname url@samestyle\endcsname
\providecommand{\newblock}{\relax}
\providecommand{\bibinfo}[2]{#2}
\providecommand{\BIBentrySTDinterwordspacing}{\spaceskip=0pt\relax}
\providecommand{\BIBentryALTinterwordstretchfactor}{4}
\providecommand{\BIBentryALTinterwordspacing}{\spaceskip=\fontdimen2\font plus
\BIBentryALTinterwordstretchfactor\fontdimen3\font minus
  \fontdimen4\font\relax}
\providecommand{\BIBforeignlanguage}[2]{{%
\expandafter\ifx\csname l@#1\endcsname\relax
\typeout{** WARNING: IEEEtran.bst: No hyphenation pattern has been}%
\typeout{** loaded for the language `#1'. Using the pattern for}%
\typeout{** the default language instead.}%
\else
\language=\csname l@#1\endcsname
\fi
#2}}
\providecommand{\BIBdecl}{\relax}
\BIBdecl

\bibitem{black05blizzard}
A.~W. Black and K.~Tokuda, ``The {B}lizzard {C}hallenge-2005: Evaluating
  corpus-based speech synthesis on common datasets,'' in \emph{Proc.
  INTERSPEECH}, Lisbon, Portugal, Sep. 2005.

\bibitem{patton16automos}
B.~Patton, Y.~Agiomyrgiannakis, M.~Terry, K.~W. Wilson, R.~A. Saurous, and
  D.~Sculley, ``{AutoMOS}: Learning a non-intrusive assessor of
  naturalness-of-speech,'' \emph{arXiv preprint arXiv:1611.09207}, 2016.

\bibitem{lo2019mosnet}
C.-C. Lo, S.-W. Fu, W.-C. Huang, X.~Wang, J.~Yamagishi, Y.~Tsao, and H.-M.
  Wang, ``{MOSNet}: Deep learning-based objective assessment for voice
  conversion,'' \emph{Proc. Interspeech}, pp. 1541--1545, 2019.

\bibitem{Leng2021MBNETMP}
Y.~Leng, X.~Tan, S.~Zhao, F.~K. Soong, X.-Y. Li, and T.~Qin, ``{MBNET}: {MOS}
  prediction for synthesized speech with mean-bias network,'' \emph{Proc.
  ICASSP}, pp. 391--395, 2021.

\bibitem{huang21voicemos}
W.-C. Huang, E.~Cooper, Y.~Tsao, H.-M. Wang, T.~Toda, and J.~Yamagishi, ``{The
  VoiceMOS Challenge 2022},'' \emph{arXiv preprint arXiv:2203.11389}, 2022.

\bibitem{cooper2021bvcc}
E.~Cooper and J.~Yamagishi, ``How do voices from past speech synthesis
  challenges compare today?'' in \emph{Proc. SSW}, 2021, pp. 183--188.

\bibitem{Hayashi2020EspnetTTSUR}
T.~Hayashi, R.~Yamamoto, K.~Inoue, T.~Yoshimura, S.~Watanabe, T.~Toda,
  K.~Takeda, Y.~Zhang, and X.~Tan, ``{ESPnet-TTS}: {U}nified, reproducible, and
  integratable open source end-to-end text-to-speech toolkit,'' \emph{Proc.
  ICASSP}, pp. 7654--7658, 2020.

\bibitem{Cooper2021GeneralizationAO}
E.~Cooper, W.-C. Huang, T.~Toda, and J.~Yamagishi, ``Generalization ability of
  {MOS} prediction networks,'' \emph{arXiv preprint arXiv:2110.02635}, 2021.

\bibitem{serra2021sesqa}
J.~Serr{\`a}, J.~Pons, and S.~Pascual, ``{SESQA}: semi-supervised learning for
  speech quality assessment,'' in \emph{Proc. ICASSP}.\hskip 1em plus 0.5em
  minus 0.4em\relax IEEE, 2021, pp. 381--385.

\bibitem{manocha2021cdpam}
P.~Manocha, Z.~Jin, R.~Zhang, and A.~Finkelstein, ``{CDPAM}: Contrastive
  learning for perceptual audio similarity,'' in \emph{Proc. ICASSP}.\hskip 1em
  plus 0.5em minus 0.4em\relax IEEE, 2021, pp. 196--200.

\bibitem{manocha2021noresqa}
P.~Manocha, B.~Xu, and A.~Kumar, ``{NORESQA}: A framework for speech quality
  assessment using non-matching references,'' \emph{Proc. NeurIPS}, vol.~34,
  2021.

\bibitem{huang21ldnet}
W.-C. Huang, E.~Cooper, J.~Yamagishi, and T.~Toda, ``{LDNet: U}nified listener
  dependent modeling in {MOS} prediction for synthetic speech,'' \emph{arXiv
  preprint arXiv:2110.09103}, 2021.

\bibitem{dbscan}
M.~Ester, H.-P. Kriegel, J.~Sander, and X.~Xu, ``A density-based algorithm for
  discovering clusters in large spatial databases with noise,'' in
  \emph{Proceedings of the Second International Conference on Knowledge
  Discovery and Data Mining}.\hskip 1em plus 0.5em minus 0.4em\relax AAAI
  Press, 1996, p. 226–231.

\bibitem{wavaugment2020}
E.~Kharitonov, M.~Rivi{\`e}re, G.~Synnaeve, L.~Wolf, P.-E. Mazar{\'e},
  M.~Douze, and E.~Dupoux, ``Data augmenting contrastive learning of speech
  representations in the time domain,'' \emph{arXiv preprint arXiv:2007.00991},
  2020.

\bibitem{wolpert1992stacked}
D.~H. Wolpert, ``Stacked generalization,'' \emph{Neural Networks}, vol.~5,
  no.~2, pp. 241--259, 1992.

\bibitem{breiman1996stacked}
L.~Breiman, ``Stacked regressions,'' \emph{Machine learning}, vol.~24, pp.
  49--64, 1996.

\bibitem{zhou2012ensemble}
Z.-H. Zhou, \emph{Ensemble methods: foundations and algorithms}.\hskip 1em plus
  0.5em minus 0.4em\relax CRC press, 2012.

\bibitem{baevski2020wav2vec}
A.~Baevski, H.~Zhou, A.~Mohamed, and M.~Auli, ``wav2vec 2.0: A framework for
  self-supervised learning of speech representations,'' \emph{arXiv preprint
  arXiv:2006.11477}, 2020.

\bibitem{librispeech}
V.~{Panayotov}, G.~{Chen}, D.~{Povey}, and S.~{Khudanpur}, ``Librispeech: {An}
  {ASR} corpus based on public domain audio books,'' in \emph{Proc. ICASSP},
  South Brisbane, Australia, Apr. 2015, pp. 5206--5210.

\bibitem{xu2021simple}
Q.~{Xu}, A.~{Baevski}, and M.~{Auli}, ``{Simple and Effective Zero-shot
  Cross-lingual Phoneme Recognition},'' \emph{arXiv preprint arXiv:2109.11680},
  2021.

\bibitem{conneau21_interspeech}
A.~Conneau, A.~Baevski, R.~Collobert, A.~Mohamed, and M.~Auli, ``{Unsupervised
  Cross-Lingual Representation Learning for Speech Recognition},'' in
  \emph{Proc. Interspeech 2021}, 2021, pp. 2426--2430.

\bibitem{ardila-etal-2020-common}
R.~Ardila, M.~Branson, K.~Davis, M.~Kohler, J.~Meyer, M.~Henretty, R.~Morais,
  L.~Saunders, F.~Tyers, and G.~Weber, ``Common voice: A massively-multilingual
  speech corpus,'' in \emph{Proc. LREC 2020}, 2020, pp. 4218--4222.

\bibitem{kingma14adam}
D.~Kingma and B.~Jimmy, ``Adam: {A} method for stochastic optimization,''
  \emph{arXiv preprint arXiv:1412.6980}, 2014.

\bibitem{akiba2019optuna}
T.~Akiba, S.~Sano, T.~Yanase, T.~Ohta, and M.~Koyama, ``Optuna: A
  next-generation hyperparameter optimization framework,'' in \emph{Proc. KDD},
  2019.

\bibitem{hsu2021hubert}
W.-N. Hsu, B.~Bolte, Y.-H.~H. Tsai, K.~Lakhotia, R.~Salakhutdinov, and
  A.~Mohamed, ``{HuBERT}: Self-supervised speech representation learning by
  masked prediction of hidden units,'' \emph{arXiv preprint arXiv:2106.07447},
  2021.

\bibitem{chen2022wavlm}
S.~Chen, C.~Wang, Z.~Chen, Y.~Wu, S.~Liu, Z.~Chen, J.~Li, N.~Kanda,
  T.~Yoshioka, X.~Xiao, J.~Wu, L.~Zhou, S.~Ren, Y.~Qian, Y.~Qian, J.~Wu,
  M.~Zeng, X.~Yu, and F.~Wei, ``{WavLM}: Large-scale self-supervised
  pre-training for full stack speech processing,'' \emph{arXiv preprint
  arXiv:2110.13900}, 2021.

\bibitem{guolin2017lightgbm}
G.~Ke, Q.~Meng, T.~Finley, T.~Wang, W.~Chen, W.~Ma, Q.~Ye, and T.-Y. Liu,
  ``{LightGBM}: A highly efficient gradient boosting decision tree,'' in
  \emph{Proc. NIPS}, vol.~30, 2017.

\end{thebibliography}

\end{document}